\documentstyle[prl,aps,multicol,epsf]{revtex}
\tighten
\begin{document}
\draft
\title{
Electron energy relaxation in the presence of magnetic impurities
}
\author{A. Kaminski and L.I. Glazman}
\address{Theoretical Physics Institute, University
of Minnesota, Minneapolis, MN 55455, USA}
\maketitle
\begin{abstract}
We study inelastic electron-electron scattering mediated by the
exchange interaction of electrons with magnetic impurities, and find
the kernel of the corresponding two-particle collision integral.  In a
wide region of parameters, the kernel $K$ is proportional to the
inverse square of the transferred energy, $K\propto J^4/E^2$. The
exchange constant $J$ is renormalized due to the Kondo effect,
yielding an additional weak dependence of $K$ on the energies of the
colliding electrons. At small energy transfers, the $1/E^2$ divergence
is cut off; the cut-off energy is determined by the dynamics of the
impurity spins. The obtained results may provide a quantitative
explanation of the experiments of Pothier {\em et al.}
[Phys. Rev. Lett. {\bf 79}, 3490 (1997)] on anomalously strong energy
relaxation in short metallic wires.
\end{abstract}

\begin{multicols}{2}

The effect of magnetic impurities on the electron properties of a
metal is drastically different from that of ``usual'' defects which
just violate the translational invariance of the crystalline lattice.
The reason for the difference is that a magnetic impurity brings an
additional degree of freedom -- its spin. If there were no
exchange interaction with the itinerant electrons, the ground state of
the system would be degenerate with respect to the orientation of
impurity spins. Weak exchange interaction allows an itinerant electron
to flip its spin in the course of scattering on a magnetic impurity.
Such scattering, accounted for even in the lowest-order (Born)
approximation, yields an important effect of dephasing of the electron
state. Finite dephasing time, in turn, suppresses the interference
corrections to the conductivity, thus suppressing the weak
localization effect\cite{WL}.

The higher-order terms in the perturbation theory series for the
scattering amplitude reveal one more important phenomenon. It turns
out that the amplitude of scattering caused by the exchange
interaction increases with lowering the temperature, as opposed to the
temperature-independent scattering on a usual impurity. This increase
is responsible for the non-monotonic temperature dependence of the
resistivity of a metal, the phenomenon called the Kondo
effect\cite{Hewson}.

Spin exchange between an electron and a magnetic impurity may occur in
an act of elastic scattering. Accounting for these spin-exchange
elastic processes is sufficient for understanding the dephasing
phenomenon\cite{WL} and the Kondo effect\cite{Kondo}. However, such
processes do not lead to any energy relaxation of electrons.  In this
paper we demonstrate that magnetic impurities may also mediate energy
transfer between electrons.  If the energy transfer $E$ is larger than
the Kondo temperature $T_K$, then the energy relaxation occurs
predominantly in two-electron collisions.  We derive the kernel $K$ of the
corresponding collision integral in the kinetic equation for the
distribution function. This kernel depends strongly on the transferred
energy, $K\propto J^4/E^2$. The dependence of $K$ on the energies
$\varepsilon_i$ of the colliding electrons (measured from the Fermi
level) is relatively weak as long as $|\varepsilon_i|\gg T_K$. This
dependence comes from the logarithmic in $|\varepsilon_i|$
renormalization of the exchange integral $J$, known from the theory of
Kondo effect\cite{Hewson}. At small energy transfers, the $1/E^2$
divergence of the kernel is cut off; the cut-off energy is determined
by the dynamics of the impurity spins, which results from their
interaction with the Fermi sea.

The motivation for our study comes from the
experiment\cite{PothierEtal97,PierreEtal00} where the relaxation of
the electron energy distribution function in mesoscopic wires was
investigated. It was found that the empirical relation $K\propto
1/E^2$ holds in a substantial interval of energies $E$ for Au and Cu
wires.  The data of Ref.~\cite{PothierEtal97,PierreEtal00} was
accurate enough to rule out the direct Coulomb
interaction\cite{AAreview}, which would yield $K(E)\propto 1/E^{3/2}$,
as a source of relaxation.

We describe the metal with magnetic impurities by means of the
exchange Hamiltonian:
\begin{eqnarray}
  \label{H}
  \hat{H}&=&\hat{H}_0+\sum_l\hat{V}_{l}\;,\;\;
\hat{H}_0\equiv\sum_{{\bf k}\alpha}\xi^{\phantom{\dagger}}_{\bf k} 
c^\dagger_{{\bf k}\alpha} c^{\phantom{\dagger}}_{{\bf k}\alpha}\;,\\
\hat{V}_{l}&\equiv&
\sum_{{\bf k}\alpha {\bf k}' \alpha'}
J e^{i({\bf k}-{\bf k}'){\bf r}_l}
\left(\hat{\bf S}_l \bbox{\sigma}_{\alpha\alpha'}\right)
c^\dagger_{{\bf k}\alpha} c^{\phantom{\dagger}}_{{\bf k}'\alpha'}\;,
\nonumber
\end{eqnarray}
where index $l$ labels the magnetic impurities, $\hat{\bf S}_l$ is
the spin operator of the $l$-th impurity, $\hat{\bf S}^2_l=S(S+1)$,
and ${\bf r}_l$ is its coordinate. Free electron states are labelled
by the wave vector ${\bf k}$ and spin index $\alpha$. The Pauli
matrices are denoted by $\bbox{\sigma}\equiv
(\sigma^x,\sigma^y,\sigma^z)$.

If the concentration $n$ of the impurities is low enough, they can be
considered independently. Therefore we will perform our calculations
for a single impurity, omitting the impurity index $l$, and then will
multiply the resulting expressions for the scattering rate by $n$.  In
this one-impurity problem, there is interaction only in $s$ channel,
so we will label the participating electron states with scalar index
$k$.

\begin{figure}
\epsfxsize=1.5in
\centerline{\epsfbox{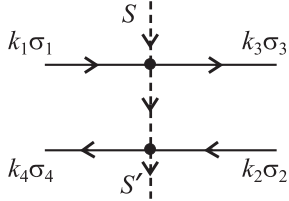}}
\caption{\label{fig1}
  A characteristic diagram for the amplitude of inelastic
  electron-electron scattering mediated by the exchange interaction of
  electrons with a magnetic impurity, in notation of
  Ref.~\protect\cite{Abrikosov}. The solid lines denote electron
  states, the dashed lines denote the localized spin state.}
\end{figure}

In the framework of the exchange Hamiltonian (\ref{H}), the lowest
non-vanishing order of the perturbation theory series in the exchange
constant $J$ for the inelastic scattering amplitude is the second
order:
\end{multicols}
\begin{eqnarray}
\label{inampt}
&&A(k_1\sigma_1,k_2\sigma_2,S \to
k_3\sigma_3,k_4\sigma_4,S')
\delta(\xi_{k_1}\!\!+\!\xi_{k_2}\!\!-\!\xi_{k_3}\!\!-\!\xi_{k_4})=
\\
&&\langle k_3\sigma_3,k_4\sigma_4,S'|\!
-\!\!\int_{-\infty}^{\infty}\!\!\! dt \!\!
\sum_{k\alpha k' \alpha'}\!\!\!
J \hat{\bf S}(t) \bbox{\sigma}_{\alpha\alpha'} 
c^\dagger_{k\alpha}(t) c^{\phantom{\dagger}}_{k'\alpha'}(t)
e^{-0|t|}
\!\!\int_{-\infty}^{t}\!\!\! dt' \!\!
\sum_{{p}\beta {p}' \beta'}\!\!\!
J \hat{\bf S}(t') \bbox{\sigma}_{\beta\beta'} 
c^\dagger_{{p}\beta}(t') c^{\phantom{\dagger}}_{{p}'\beta'}(t')
e^{-0|t'|}
|k_1\sigma_1,k_2\sigma_2,S\rangle\;.\nonumber
\end{eqnarray}
\begin{multicols}{2}
In the diagrammatic representation, the amplitude is the sum of the
diagram shown in Fig.~\ref{fig1} and the three other diagrams that can
be obtained from the diagram of Fig.~\ref{fig1} by the transposition
of indices $1\leftrightarrow 2$ and/or $3\leftrightarrow 4$.  Note
that there is no summation over the initial or final spin states of
the impurity in Eq.~(\ref{inampt}). Therefore, the spin lines are not
closed, {\em i.e.} contrary to Ref.~\cite{SolyomZawadowski69} this
scattering amplitude cannot be represented in the  form of an effective
four-electron vortex.

Performing the time integrations in Eq.~(\ref{inampt}), we obtain the
standard expression for the second-order term of the expansion of
$T$-matrix,
\begin{eqnarray} 
\label{inamp} 
&&A(k_1\sigma_1,k_2\sigma_2,S
\to k_3\sigma_3,k_4\sigma_4,S')=\\ 
&&\langle k_3\sigma_3,k_4\sigma_4,S'| 
\hat{V}\frac{1}{\xi_{k_1}\!\!+\!\xi_{k_2}-\hat{H}_0}
\hat{V} |k_1\sigma_1,k_2\sigma_2,S\rangle\;.
\nonumber 
\end{eqnarray}                                                                  
The denominator in Eq.~(\ref{inamp}) is the energy of the intermediate
virtual state, which equals $\pm(\xi_{k_1}-\xi_{k_3})$ for two of
the four possible pairings of the electron creation-annihilation
operators in Eqs.~(\ref{inampt}) and (\ref{inamp}) (one of these pairings is
shown on Fig.~\ref{fig1}), or $\pm(\xi_{k_1}-\xi_{k_4})$ for the
other two pairings. The spin structure of the scattering amplitude can
easily be found from Eq.~(\ref{inamp}). In a scattering event, spins
of one or both participating electrons must flip, with the
corresponding change of the impurity spin. In this paper we are
interested only in the relaxation of the electron energy distribution,
and assume that the system does not have any spin polarization.
Therefore we need to calculate only the total cross-section of
scattering into all possible spin states, averaged over the initial
spin states of the impurity and two electrons. After this averaging,
the terms proportional to
$[(\xi_{k_1}-\xi_{k_3})(\xi_{k_1}-\xi_{k_4})]^{-1}$  
drop out. Finally, we get the rate of
scattering of two electrons with energies $\varepsilon_1$ and
$\varepsilon_2$ into states with energies $\varepsilon_3$ and
$\varepsilon_4$:
\begin{eqnarray}
  \label{inrate4}
&&  \Gamma(\varepsilon_1,\varepsilon_2;\varepsilon_3,\varepsilon_4)
=\frac{\pi}{4} \frac{n}{\nu} S(S+1) (J\nu)^4 \\
&&\;\;\;\times\left[\frac{1}{(\varepsilon_1-\varepsilon_3)^2}
+\frac{1}{(\varepsilon_1-\varepsilon_4)^2}\right]
\delta(\varepsilon_1+\varepsilon_2-\varepsilon_3-\varepsilon_4)
\;.\nonumber
\end{eqnarray}
Substituting this expression into the collision
integral\cite{AbrikosovBook} and taking into account the symmetry
$\varepsilon_1\leftrightarrow\varepsilon_2$,
$\varepsilon_3\leftrightarrow\varepsilon_4$ of the right-hand side of
Eq.~(\ref{inrate4}), we obtain the collision integral in the form:
\begin{eqnarray}
  \label{relax}
I\left(\varepsilon,\left\{f\right\}\right)
&=&\int\!\!\int dE d\varepsilon' 
  K(E)\\
&\times&\left\{f(\varepsilon)f(\varepsilon')\left[1-f(\varepsilon-E)\right]
\left[1-f(\varepsilon'+E)\right]\right.\nonumber\\
&-&\left.f(\varepsilon-E)f(\varepsilon'+E) 
\left[1-f(\varepsilon)\right]\left[1-f(\varepsilon')\right]\right\}\;,
\nonumber
\end{eqnarray}
where the kernel
\begin{equation}
  \label{KE}
K(E)= \frac{\pi}{2}\frac{n}{\nu}S(S+1)(J\nu)^4
\frac{1}{E^2}
\end{equation}
depends only on the energy $E$ transferred in the
collision, and $f(\varepsilon)$ is the electron
energy distribution function.

The above derivation of the inelastic amplitude [Eq.~(\ref{inamp})]
was performed in the lowest non-vanishing order of the perturbation
theory. It is known from the theory of the Kondo effect, that for the
elastic scattering amplitude, the calculation in the lowest order may
be insufficient. The higher in $J\nu$ orders yield contributions to
the elastic scattering which logarithmically diverge at low
energies\cite{Kondo,Abrikosov}. For elastic scattering, the
leading terms in all orders can be summed up with the help of the
renormalization-group technique\cite{Anderson}.  In this technique,
the bare exchange constant is replaced by the renormalized one,
\begin{equation}
  \label{Jrenorm}
  J(\varepsilon)=\left[\nu \ln (|\varepsilon|/T_K)\right]^{-1}\;,
\end{equation}
and the scattering amplitude at energy $\varepsilon$ is to be
calculated within the Born approximation in $J(\varepsilon)$.  The
Kondo temperature $T_K$ is related to the parameters of the
Hamiltonian (\ref{H}) by
\begin{equation}
  \label{TK}
  T_K=\mu\sqrt{J\nu}D\exp\left(-1/J\nu\right)\;.
\end{equation}
Here $D$ is the high-energy cut-off, and $\mu\sim 1$ (for detailed
discussion of these parameters, see Ref.~\cite{Haldane}). 

Similar to the theory of elastic scattering, the lowest-order result
(\ref{KE}) is valid only while $\ln(|\varepsilon_i|/T_K)\gg 1$. At smaller
energies, the two vertices of Fig.~\ref{fig1} acquire corrections of
the form $J\nu[J\nu\ln (D/|\varepsilon_i|)]^m$ in the $(m+1)$--st
order of the perturbation theory.  The two vertices are renormalized
independently from each other\cite{otherdiags}, and the corresponding
diagrams have the same structure as in Ref.\cite{Abrikosov}.  This
approximation is justified as long as the energies $\varepsilon_i$ of
all incoming and outgoing electrons satisfy the condition
\begin{equation}
\label{logcond}
\ln\left(|\varepsilon_i|/T_K\right)\gtrsim 1\;.
\end{equation}
The resulting non-perturbative expression for the kernel in the collision
integral reads
\begin{eqnarray}
K(E;\varepsilon,\varepsilon')&=& 8\pi\frac{n}{\nu} S(S+1)
{\left[\ln\displaystyle\frac{|\varepsilon|}{T_K}+
\ln\displaystyle\frac{|\varepsilon-E|}{T_K}\right]^{-2}}\nonumber\\
&\times&
{\left[\ln\displaystyle\frac{|\varepsilon'|}{T_K}
+\ln\displaystyle\frac{|\varepsilon+E|}{T_K}\right]^{-2}}
\frac{1}{E^2}\;.
  \label{KERG}
\end{eqnarray}
We would like to emphasize that the dependence of the kernel $K$ on
the energies of participating electrons remains weak (logarithmic), as
opposed to the strong $1/E^2$ dependence on the transferred energy
$E$.  The logarithmic dependence in Eq.~(\ref{KERG}) is meaningful as
long as the electron energies $\varepsilon_i$ exceed the smearing of
the Fermi level caused by temperature or a non-equilibrium electron
distribution. In the opposite case, the arguments of all the
logarithms in Eq.~(\ref{KERG}) must be replaced by
$\varepsilon^*/T_K$, where energy $\varepsilon^*$ characterizes the
smearing. It is important to note that $\varepsilon^*$ does not cut
off the singularity in the transferred energy $E$.

The low-energy divergence of the scattering rate (\ref{inrate4}) stems
from the degeneracy of spin states of the impurity. Due to this
degeneracy, the energy of the intermediate virtual state [the
denominator in Eq.~(\ref{inamp})] approaches zero at $E\to 0$.  An
additional condition for the divergency is the time independence of
the average $\langle S'| \hat{S}^{j}(t) \hat{S}^{k}(t')| S\rangle$ in
the approximation of Eq.~(\ref{inampt}) (here $j,k=x,y,z$). In fact,
exchange interaction between the itinerant electrons and impurity may
flip its spin.  The resulting impurity spin correlation function
decays, restricting the lifetime of the intermediate virtual state.
The corresponding decay rate cuts off the $E=0$ singularity of the
kernel (\ref{KERG}). The manner of decay depends on the electron
energy distribution $f(\varepsilon)$. We will discuss first the
cut-off in the case of weak deviations of $f(\varepsilon)$ from the
thermal equilibrium.

Let us first consider the low-energy cut-off for $K(E)$ at high
temperatures, $T\gg T_K$.  Scattering of electrons off the spin
results in exponential decay of the correlation function, $\langle
S'| \hat{S}^{j}(t) \hat{S}^{k}(t')| S\rangle\propto
\exp(-|t-t'|/\tau_T)$.  The impurity spin correlation time $\tau_T$ can be
evaluated with the help of the Fermi golden rule. Since the deviation
from the thermal equilibrium is weak, we can replace $f(\varepsilon)$
with the Fermi distribution function $n_F(\varepsilon)$,
\begin{eqnarray}
\frac{\hbar}{\tau_{T}}&=&\frac{2\pi}{3}S(S+1) (J\nu)^2 
\int d\varepsilon n_F(\varepsilon)[1-n_F(\varepsilon)]\nonumber\\
&=&\frac{2\pi}{3} S(S+1) (J\nu)^2 T\;.
\label{spin-flip-born}
\end{eqnarray}
 When $T$ is lowered towards $T_K$, the
exchange constant in Eq.~(\ref{spin-flip-born}) is renormalized
according to Eq.~(\ref{Jrenorm}) with $\varepsilon=T$. The resulting
expression for the spin-flip rate reads:
\begin{equation}
\frac{\hbar}{\tau_{T}}=\frac{2\pi}{3} S(S+1)
\left[\ln (T/T_K)\right]^{-2} T\;. 
\label{spin-flip-bornRG}
\end{equation}
The energy scale $\hbar/\tau_T$ sets the limit of applicability of
Eq.~(\ref{inamp}) and cuts off the singularity in the kernel (\ref{KERG}) at
$E\sim \hbar/\tau_T$. Note that within the limits of applicability of
Eq.~(\ref{spin-flip-bornRG}), the spin-flip rate satisfies the
condition $T>\hbar/\tau_T>T_K$.

At $T< T_K$, we can neglect the effect of a finite temperature on the
scattering of electrons with $|\varepsilon_i|\gg T_K$. The behavior of
$K(E)$ can be related to the time dependence of the zero-temperature
self-correlation function of the impurity spin. At time difference
$|t-t'|\gg \hbar/T_K$ this correlation function decays as $\langle GS|
\hat{S}^{j}(t) \hat{S}^{k}(t')|GS\rangle\sim \delta_{jk}/[T_K
|t-t'|]^2$, and therefore the singularity of the kernel $K(E)$ is cut
off at $E\sim T_K$ (here $|GS\rangle$ is the ground state of the Kondo
problem).

At this point, we should mention that at energy transfers $E\sim T_K$,
the processes with participation of three or more electrons must also
be taken into account along with the two-particle scattering. The
consideration of these multi-particle processes is an arduous task
lying beyond the scope of this paper. Here we address mostly the
electron energy relaxation on large ($\gtrsim T_K$) energy scales.  It
allows us to limit our consideration to the two-particle processes
accounted for by Eq.~(\ref{relax}), and dispense with the scattering
events involving more particles.

At very small energies ($|\varepsilon_i|,T\ll T_K$), however, the
Fermi-liquid description of electrons is again a valid tool. The
behavior of the system is described in this case by the quadratic
fixed-point Hamiltonian, in which the four-fermion interaction is a
least-irrelevant term\cite{Nozieres,AffleckLudwig93}. The calculation
of the inelastic scattering rate is then straightforward, the
resulting collision-integral kernel is given by
\begin{equation}
  \label{KENozieres}
  K(E)=\frac{1}{T_K^2}\frac{n}{\nu}\;.
\end{equation}
When $T=0$, the corresponding rate of inelastic electron scattering is
$\hbar/\tau_{\rm in}=\int_0^\varepsilon dE K(E) E\propto
(\varepsilon/T_K)^2$.  At $\varepsilon\to 0$, it decreases faster than
$\varepsilon$, as it is supposed to be in the Fermi-liquid picture.

Relaxation of the electron energy distribution was investigated
experimentally in metallic wires of Cu and Au in
Refs.~\cite{PothierEtal97,PierreEtal00}. In these
experiments, a finite bias $V=50-500$ $\mu$eV was applied to
the ends of a wire. It was found that starting from
fairly small wire lengths, the electron distribution is smeared over
the range of energies $eV$, instead of having two distinct steps
created by the bias applied to the wire ends.  The observed electron
energy relaxation was attributed~\cite{PothierEtal97,PierreEtal00} to
two-electron collisions.
The collision-integral kernel for $E<eV$ extracted from the
experiments has the form $K(E)=\hbar/(\tau_0 E^2)$, with a cut-off at
some low energy, which scales linearly with $eV$ \cite{private}.  The
value of the parameter $\tau_0$ was $0.5-1.0$ ns for Cu wires,
and $0.1$ ns for Au wires~\cite{GougamEtal00}.

Now we discuss the possibility of such relaxation due to the electron
scattering on magnetic impurities in wires.  In the experimental
setup, the electron distribution is smeared, and the width of smearing
$eV$ exceeds both $T$ and the energies $|\varepsilon_i|$ of the
colliding electrons.  Assuming also that $eV\gg T_K$, we can simplify
the kernel (\ref{KERG}):
\begin{equation}
  \label{KEV}
K(E)= \frac{\pi}{2}\frac{n}{\nu}S(S+1)[\ln(eV/T_K)]^{-4}
\frac{1}{E^2}\;.
\end{equation}
The $1/E^2$ dependence in Eq.~(\ref{KEV}) persists down to the
cut-off, which is determined by the spin-flip rate $1/\tau_{eV}$.  The
derivation of the spin-flip rate in these non-equilibrium conditions
follows the same path which led to Eq.~(\ref{spin-flip-bornRG}), and
results in:
\begin{equation}
\label{spin-flipRG}
\frac{\hbar}{\tau_{eV}}=\gamma S(S+1)[\ln(eV/T_K)]^{-2} eV.
\end{equation}
Here the numerical constant $\gamma\sim 1$ depends on the detailed
shape of the non-equilibrium electron distribution. 

Now we compare theoretical results (\ref{KEV}) and (\ref{spin-flipRG})
with the experimental observations for Au wires~\cite{PierreEtal00}.
Properties of these samples are compatible with the presence of iron
impurities with a concentration up to few tens of ppm~\cite{private}.
We take the density of states in Au at the Fermi level $\nu\approx
0.25\ (\mbox{eV}\ \mbox{site})^{-1}$~\cite{Kittel}, and $T_K\approx
0.3$K for Fe impurities in Au~\cite{Hewson}. The typical value of
voltage in the experiments \cite{PierreEtal00} was $V\approx 0.3$
meV~\cite{private}.  Substituting these parameters into
Eq.~(\ref{KEV}), we see that a relatively small concentration $n\sim
10$ ppm is sufficient to reproduce the experimentally measured value
$\tau_0\approx 0.1$ ns\cite{tauphi}.  The spin-flip rate
(\ref{spin-flipRG}) is the low-energy cut-off for the $1/E^2$
dependence of the kernel.  This cut-off is roughly proportional to the
applied voltage, in agreement with experimental
observations~\cite{private}. We must note, however, that the lower
voltages used in experiment~\cite{PierreEtal00} are close to the Kondo
temperature, so the leading-logarithmic
approximation~\cite{Abrikosov,Anderson}, used in derivation of
Eqs.~(\ref{KEV})--(\ref{spin-flipRG}), may be insufficient.

The inelastic electron scattering off magnetic impurities
must be sensitive to an external magnetic field polarizing the spins
in the system. The Zeeman splitting $g\mu H$ of the spin states
prevents the impurity spins from changing, thus suppressing the
inelastic scattering processes with energy transfers $E<g\mu H$.
Measurements in a sufficiently strong magnetic field may elucidate the
role of magnetic impurities in the electron energy relaxation. 

An important feature of the electron-electron interaction mediated by
magnetic impurities is that it is not translationally invariant.  This
is why the introduction of non-magnetic impurities, which affects
drastically Coulomb interaction of electrons\cite{AAreview}, produces
only small corrections to the interaction induced by magnetic
impurities.

The above consideration was performed for the magnetic impurities
described by the one-channel Kondo model. However, the proportionally
of the scattering integral kernel to $1/E^2$ at $E\gg T_K$ holds for
an arbitrary number $N$ of channels in the Kondo problem. If 
the exchange constants $J$ are the same for all channels, the kernel
(\protect\ref{KE}) acquires an additional factor $N^2$. For the
specific case $N=2$, the qualitative $K(E)\propto 1/E^2$ behavior was
noticed in Ref.~\cite{multichannel}. However in Ref.~\cite{multichannel}
 this behavior was
attributed {\it solely} to $N=2$; this, in our opinion, is
inaccurate. Our consideration also indicates that at any $N$, the
$1/E^2$ divergence of the kernel $K(E)$ is cut-off at small $E$, see
Eq.~(\ref{spin-flipRG}). This is also in
an apparent disagreement with Ref.~\cite{multichannel}, which states
that for $N=2$ the divergence persists down to the lowest energies.

In conclusion, we have shown that the exchange interaction of
itinerant electrons with magnetic impurities can facilitate inelastic
electron-electron scattering. We derived the kernel of the
corresponding collision integral, and found its explicit dependence on
the parameters of the system for a wide range of the energies of
colliding electrons. This allowed us to perform a quantitative
analysis of the experimental results of
Refs.~\cite{PothierEtal97,PierreEtal00}. We find that a very small
density of magnetic impurities could lead to the anomalies in the
electron energy relaxation observed there. 

This work was supported by NSF Grants No. DMR 97-31756 and DMR
98-12340. A. K. is a recipient of University of Minnesota's Doctoral
Dissertation Fellowship. The authors are grateful to I. Aleiner, N.
Birge, D. Esteve, A. Ludwig and H. Pothier for valuable discussions.

\end{multicols}

\begin{references}
\bibitem{WL} B.L. Altshuler {\em et al.}, Sov. Phys. JETP {\bf 54},
  411 (1980). 
\bibitem{Hewson} A.C. Hewson, {\em The Kondo problem to heavy
fermions} (Cambridge University Press, 1993).
\bibitem{Kondo} J. Kondo, Prog. Theor. Phys. {\bf 32}. 37 (1964).
\bibitem{PothierEtal97} H. Pothier {\em et al.},
Phys. Rev. Lett. {\bf 79}, 3490 (1997).
\bibitem{PierreEtal00} F. Pierre {\em et al.}, cond-mat/0012038 and
  in: {\em Proceedings of  the NATO Advanced Research Workshop on
  size-dependent magnetic scattering}, Pecs, Hungary, 2000, in press. 
\bibitem{AAreview} B.L. Altshuler, A.G. Aronov, in {\em Electron-electron
  interactions in disordered systems}, ed. by A.L. Efros and M. Pollak
(North-Holland, Amsterdam, 1985).
\bibitem{SolyomZawadowski69} J. S\'{o}lyom, A. Zawadowski, Z. Physik
{\bf 226}, 116 (1969).
\bibitem{Abrikosov} A.A. Abrikosov, Physics {\bf 2}, 21 (1965).
\bibitem{AbrikosovBook} A.A. Abrikosov, {\em Fundamentals of the
  theory of metals} (New York, Elsevier, 1988), p. 57.
\bibitem{Anderson} P.W. Anderson, J. Phys. C {\bf 3}, 2436 (1970).
\bibitem{Haldane} F.D.M. Haldane, J. Phys. C {\bf 11}, 5015 (1978).
\bibitem{otherdiags}Diagrams with more than one spin line connecting
the two electron lines of Fig.~\protect\ref{fig1} either yield zero after
summation over the spin indices of intermediate states, or
are proportional to a lower power of the logarithmic factor.
\bibitem{Nozieres} P. Nozi\`{e}res, J. Low Temp. Phys {\bf 17}, 31
(1974).
\bibitem{AffleckLudwig93} I. Affleck and A.W.W. Ludwig, Phys. Rev. B
{\bf 48}, 7297 (1993).
\bibitem{private}H. Pothier, private communication.
\bibitem{GougamEtal00}A.B. Gougam {\em et al.}, 
J. Low Temp. Phys. {\bf 118}, 447 (2000).
\bibitem{Kittel} C. Kittel, {\em Introduction to solid state physics}
(John Wiley \& sons, New York, 1996). 
\bibitem{tauphi} In the framework of our model, the electron dephasing
  time is given by $\tau_\phi\sim\tau_0/\ln^2(eV/T_K)$. The value $n\sim
  10$ ppm yields the estimate $\tau_\phi\sim 10$ ps, in agreement with
  the experiments\protect\cite{PierreEtal00}. 
\bibitem{multichannel}  J. Kroha, Adv. Solid State Phys. {\bf 40}, 267
  (2000).  
\end{references}
\end{document}